\documentclass[apjl]{emulateapj}

\def\ag{ }

\def\be{\begin{equation}}
\def\en{\end{equation}}

\shortauthors{Giuliani et al.}
\usepackage{subfigure}
\begin{document}

\title{Neutral pion emission from accelerated protons in the Supernova Remnant W44}

\author{A.~Giuliani\altaffilmark{1}, M.~Cardillo \altaffilmark{2,3}, M.~Tavani\altaffilmark{2,4,5}, Y.~Fukui\altaffilmark{6}, S.~Yoshiike\altaffilmark{6},
K.~Torii\altaffilmark{6}, \\G.~Dubner\altaffilmark{7}, G.~Castelletti\altaffilmark{7}, 
G.~Barbiellini\altaffilmark{8}, A.~Bulgarelli\altaffilmark{9},
P.~Caraveo\altaffilmark{1}, E.~Costa\altaffilmark{2}, P.W.~Cattaneo\altaffilmark{10}, A.~Chen\altaffilmark{1}, T.~Contessi\altaffilmark{1},
E.~Del Monte\altaffilmark{2}, I.~Donnarumma\altaffilmark{2}, Y.~Evangelista\altaffilmark{2}, M.~Feroci\altaffilmark{2}, F.~Gianotti\altaffilmark{9},
F.~Lazzarotto\altaffilmark{2}, F.~Lucarelli\altaffilmark{12}, F.~Longo\altaffilmark{8}, M.~Marisaldi\altaffilmark{9}, S.~Mereghetti\altaffilmark{1},
L.~Pacciani\altaffilmark{2}, A.~Pellizzoni\altaffilmark{11}, G.~Piano\altaffilmark{2}, P.~Picozza\altaffilmark{3}, C.~Pittori\altaffilmark{12}, 
G.~Pucella\altaffilmark{13}, M.~Rapisarda\altaffilmark{2}, A.~Rappoldi\altaffilmark{10}, S.~Sabatini\altaffilmark{2}, P.~Soffitta\altaffilmark{2}, E.~Striani\altaffilmark{2},
M.~Trifoglio\altaffilmark{9}, A.~Trois\altaffilmark{11}, S.~Vercellone\altaffilmark{14}, F.~Verrecchia\altaffilmark{12}, V.~Vittorini\altaffilmark{2},
\\S.~Colafrancesco\altaffilmark{16,17},  P.~Giommi\altaffilmark{12}, G.~Bignami\altaffilmark{15}}

\altaffiltext{1} {INAF-IASF Milano, via E.Bassini 15, 20133 Milano, Italy}
\altaffiltext{2} {INAF/IASF-Roma,via Del Fosso del Cavaliere 100, 00133 Roma, Italy}
\altaffiltext{3} {Dipartimento di Fisica, Univ.di Roma ``Tor Vergata'', via della Ricerca Scientifica 1, 00133 Roma, Italy}
\altaffiltext{4}{Consorzio Interuniversitario Fisica Spaziale (CIFS), villa Gualino, v.la Settimo Severo 63, 10133, Torino, Italy}
\altaffiltext{5} {INFN Roma Tor Vergata, via della Ricerca Scientifica 1, 00133 Roma, Italy}
\altaffiltext{6} {Department of Astrophysics, Nagoya University, Chikusa-ku, Nagoya 464-8602, Japan}
\altaffiltext{7} {Instituto de Astronomia y Fisica del Espacio (IAFE, CONICET-UBA), 1428 Buenos Aires, Argentina}
\altaffiltext{8} {Dipartimento di Fisica and INFN, Via Valerio 2, I-34127 Trieste, Italy}
\altaffiltext{9} {INAF/IASF-Bologna, Via Gobetti 101, I-40129 Bologna, Italy}
\altaffiltext{10} {INFN-Pavia, Via Bassi 6, I-27100 Pavia, Italy}
\altaffiltext {11} {INAF, Osservatorio Astronomico di Cagliari, Poggio dei Pini, strada 54, i-09012 Capoterra, italy}
\altaffiltext{12} {ASI-ASDC, Via G.Galilei, I-00044 Frascati (Roma), Italy}
\altaffiltext{13} {ENEA-Frascati, Via E.Fermi 45, I-00044 Frascati (Roma), Italy}
\altaffiltext{14} {INAF/IASF-Palermo, Via U.La Malfa 153, I-90146 Palermo, Italy}
\altaffiltext{15} {Istituto Universitario di Studi Superiori (IUSS), viale Lungo Ticino Sforza 56, 27100 Pavia, Italy}
\altaffiltext{16} {INAF, Osservatorio Astronomico di Roma, Monte Porzio Catone, Italy}
\altaffiltext{17} {School of Physics, University of the Witwatersrand, Johannesburg Wits 2050, South Africa.}

\begin{abstract}
We present the \ag{AGILE} gamma-ray observations in the energy range 50~MeV - 10~GeV of the supernova remnant (SNR) W44, 
one of the most interesting systems for studying cosmic-ray production.
W44 is an intermediate-age SNR ($\sim20,000$~years) and its ejecta expand in a dense medium as shown by a prominent radio shell, nearby molecular clouds, and bright [SII] emitting regions. 
We extend our gamma-ray analysis to energies substantially lower than previous measurements which could not conclusively establish the nature of the radiation. 
We find that gamma-ray emission matches remarkably well both the position and shape of the inner SNR shocked plasma. 
Furthermore, the gamma-ray spectrum shows a prominent peak near 1~GeV with a clear decrement at
energies below a few hundreds of MeV as expected from neutral pion decay. 
Here we demonstrate that: 
(1) hadron-dominated models are consistent with all W44 multiwavelength constraints derived from radio, optical, X-ray, and gamma-ray  observations; 
(2) \ag{{ad hoc}  lepton-dominated models fail} to explain simultaneously the well-constrained gamma-ray and radio spectra, and require a circumstellar
density much larger than the value derived from observations; \ag{(3) the hadron energy spectrum is well described by a power-law
(with index $s=3.0 \pm 0.1 $) and a}\ag{low-energy cut-off at $E_{c}= 6\pm 1$~GeV. 
Direct evidence for pion emission is then established in an SNR for the first time.}
\end{abstract}


\maketitle

\section{Introduction}
Providing an unambiguous proof of the cosmic-ray origin until now has been elusive, despite many decades of attempts and controversial claims
\citep[ {e.g.},][]{fermi,ginzburg,torres,aharoniana,berezhko,butt}.
Cosmic-rays are mainly protons and heavy ions (hadrons) and, in a few percent, electrons and positrons.
Supernova remnants (SNRs) are ideal candidates for the cosmic-ray production up to energies near $E_{knee}$ = $10^{15}$~eV.
{The SNR energy output in the Galaxy can indeed supply the energy budget necessary to maintain the present population of cosmic-rays.
Furthermore the observations of  ultra-relativistic electrons support the hypothesis that also protons are accelerated in these objects
\citep[for a recent review see][and references therein]{reynolds}.}
Proving the fact that the SNR origin of hadronic cosmic-rays is difficult because of the complexity of the SNR-environment interaction.
{From an observational point of view, a direct proof can be given by an unambiguous detection of the gamma-ray emission expected from neutral pion decay in hadronic interactions.
However radiation from co-spatially accelerated electrons can mask and sometimes overcome the expected neutral pion decay signature of proton/ion emission in the 100~MeV-a few~TeV energy range.}
Recent analysis suggests that several gamma-ray observations of SNRs can be understood in terms of accelerated hadrons
{\citep{castro10,giuliani, abdo10IC443,abdo10W28,abdo10W44,abdo09}}. 
{However it is currently not possible to exclude that the observed $\gamma$-ray emission is produced by leptons alone.
The discrepancies between leptonic and hadronic models are expected to be more evident at low energies (50 - 100~MeV). 
Gamma-ray astronomy in this energy band is very challenging because of high background-noise flux and of the strong multiple scattering suffered by electrons originating from $\gamma$-ray events. 
The AGILE/GRID instrument (calibrated in the  50~MeV - 10~GeV band), however, has already shown its ability to provide an energy spectrum starting at 50~MeV for
bright objects \citep{vercellone09,giulianiGRB,vittorini11}.
In this paper we report on a low-energy $\gamma$-ray and multiwavelength spectrum for the SNR W44 in order to constraint the emitting particle spectrum and discriminate between leptonic and hadronic models.}
\begin{figure*}[ht!]
 \begin{center}
\includegraphics[scale=0.4]{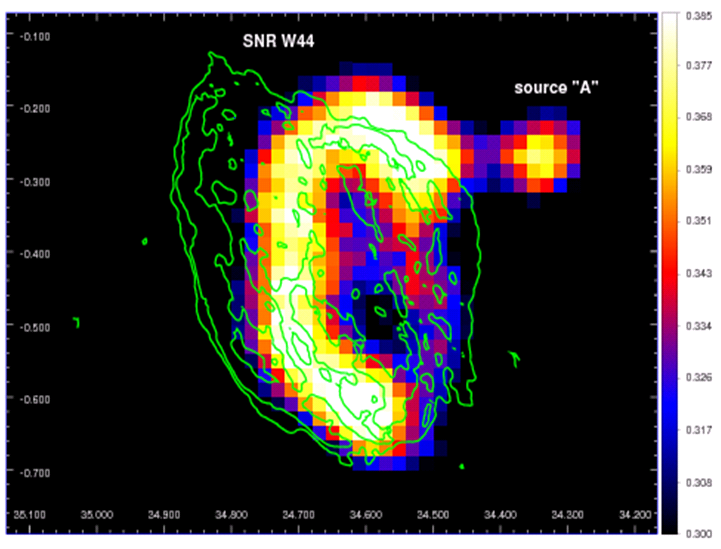}
\end{center}
\centering
\subfigure{\includegraphics[bb=0 0 500 367,scale=0.4]{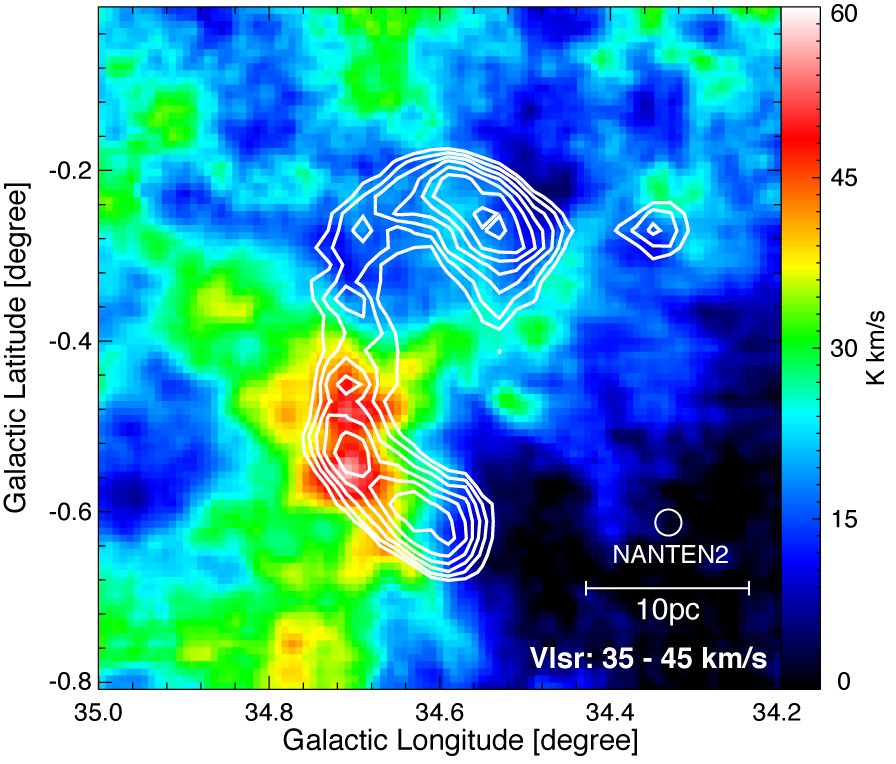}}\qquad
\subfigure{\includegraphics[bb=0 0 262 260,scale=0.4]{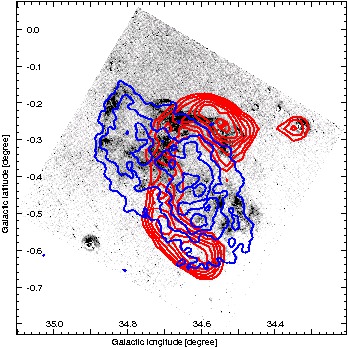}}

\caption[width=.4\textwidth]{\textbf{(a)}: AGILE gamma-ray intensity map (in Galactic coordinates) of the W44 region (1.1 x 0.75 degrees)
in the energy range 400~MeV - 3~GeV obtained by integrating all available data collected during the period 2007 July and 2011 April. The color bar scale
is in units of $10^{-5}$~photons~cm$^{-2}$~s$^{-1}$~pixel$^{-1}$.
Pixel size is 0.02 degrees with a six-bin Gaussian smoothing. Green contours show the 324 MHz radio continuum flux density detected by the Very Large Array.
 {The white cross indicates the position of the pulsar PSR B1803+01}. Source A does not appear to be associated with W44. 
\textbf{(b)}: combined CO data from the NANTEN Observatory superimposed with the AGILE gamma-ray data contours above 400~MeV (in white) of
the W44 region (map in Galactic coordinates). CO data have been selected in the velocity range 35 - 45~km~s$^{-1}$ corresponding to a kinematic distance
compatible with the W44 distance. \textbf{(c)}: the [SII] image of SNR W44 obtained in 1993 with the 0.6 m Burrell-Schmidt telescope at Kitt
Peak National Observatory with a matched red continuum image subtracted to emphasize the faint nebulosity  \citep[adapted from][]{giacani}. 
The overlaid contours trace the gamma-rays emission detected by AGILE between 400~MeV and 3~GeV (in red) and the X-ray emission detected by ROSAT in the
0.1 - 2.4~keV range (in blue).}
\label{w44}
\end{figure*}

\section{The Supernova Remnant W44}

{SNR W44 \ag{(G34.7 $ -$0.4)} is a well studied middle-aged\ag{($\sim$20,000~yr)} SNR located in the Galactic disk at a distance of  $\sim3$~kpc from Earth {\citep{clark76,wolszczan}}. 
W44 is an ideal system to test the presence of accelerated hadrons and the interplay between hadronic and leptonic models.
Radio \citep[][ and references therein]{castelletti07} and X-ray \citep{Watson} mapping of the SNR show a roughly elliptical shocked shell and a centrally peaked emission respectively. 
In the IR band \citep{reach} it is evident that the shell is expanding into a dense surrounding medium ($n\sim100$~cm$^{-3}$).
\cite{wolszczan} discovered the radio pulsar PSR~B1853+01 with distance and age compatible with the SNR.
\cite{wootten} and then \cite{rho} found a massive molecular cloud interacting with the south-eastern side of the remnant.
Evidence for a more complex system of massive MCs and for their interactions with the remnant, shown by some features typical of a strong shock, was given by \cite{seta} and then by \cite{reach}.
The MC-SNR interactions were confirmed by the maser  OH (1720~MHz) emission reported by  {\cite{claussen97}} and then by \cite{hoffman}.\\
The first estimation of the spectral radio index variations as a function of position over the remnant was done by \cite{castelletti07}:  the eastern limb spectrum was consistent with a diffusive shock acceleration model and the spectrum flattening in the westernmost arc confirmed the MC-SNR
interaction observed in IR and optical band.\\
Gamma-ray emission from this SNR has been detected by the Fermi/LAT instrument at energy E$>$200~MeV \citep{abdo10W44},
suggesting the presence of accelerated
protons interacting with the surrounding medium.}

\section{Data Analysis}
 {AGILE-GRID data were analyzed using the AGILE Standard Analysis Pipeline. We used $\gamma$-ray events filtered by means of
the $FM3.119\_2$ AGILE Filter Pipeline \citep[as described in][]{vercellone08}. 
In order to discriminate between background events and gamma rays, the GRID and anticoincidence system (ACS) signals are processed,
reconstructed and selected by a dedicated software \citep{giuliani06}.
We used the most recent versions of the diffusion model \citep{giuliani04} and of the calibration files, available at the ASDC site
(www.asdc.asi.it). We created counts, exposure and Galactic background gamma-ray maps with a bin-size of $0.02^{\circ}$ x $0.02^{\circ}$.
In order to derive the source average flux and spectrum we ran the AGILE point source analysis software ALIKE
\citep[][based on the maximum likelihood technique described in \cite{mattox93}]{bulgarelli11}
over the whole observing period 2007 July - 2011 April. 
Both statistic and systematic uncertainties are taken into account.
The spectrum was obtained by computing the $\gamma$-ray flux in six energy bins selected with the aim to have a significance $\sigma > 4$. 
In order to study the source morphology, we obtained an intensity map integrated over the energy range where the GRID angular resolution is optimal (E$>$400~MeV).}

\section{Results and discussion}
\label{results}

AGILE detects SNR W44 with a significance of 15.8 $\sigma$ as an extended source. 
Figure~\ref{w44}a shows AGILE gamma-ray intensity map above 400~MeV of the W44 region with the 324~MHz VLA radio contours.
The gamma-ray morphology remarkably resembles the quasi-elliptical pattern of the interior of the radio shell, especially coinciding with the radio brightness enhancements toward the north-west and south-east regions.
{Both the radio pulsar PSR B1803+01 position \citep{petre02} and its (small) pulsar wind nebula \citep[][and references therein]{giacani} are inconsistent with
the gamma-ray morphology detected by AGILE.}
Moreover, \cite{abdo10W44} excluded the presence of a pulsation in the  $\gamma$-ray signal.
Figure~\ref{w44}b shows the CO emission at a kinematic velocity compatible with the distance of W44, tracing the presence of MCs in the W44 surroundings, together with the gamma-ray contour levels.
It can be inferred that the south-eastern side of the $\gamma$-ray source overlaps with the MC - SNR interaction region.
In the northern part of the shell, the $\gamma$-ray and CO emissions are not correlated, however many studies of the surrounding interstellar medium showed the presence of dense gas not traced by CO \citep{reach}. 
Figure~\ref{w44}c displays an SII map overlaid with X-ray and gamma-ray contours.
{The strong sulfur [SII] emission, along with H$\alpha$ emission, indicates the presence of shocked gas \citep{draine}
.}
\footnote{{As pointed out  by \cite{rho} in their X-ray and optical study of W44, the optical filaments and the X-ray image showing locally bright
emission clumps along the filaments suggest that both are produced by the interaction between the SN shock front and regions of enhanced ambient density.}}

\begin{figure}[h!]
\begin{center}
\includegraphics[width=9.3cm]{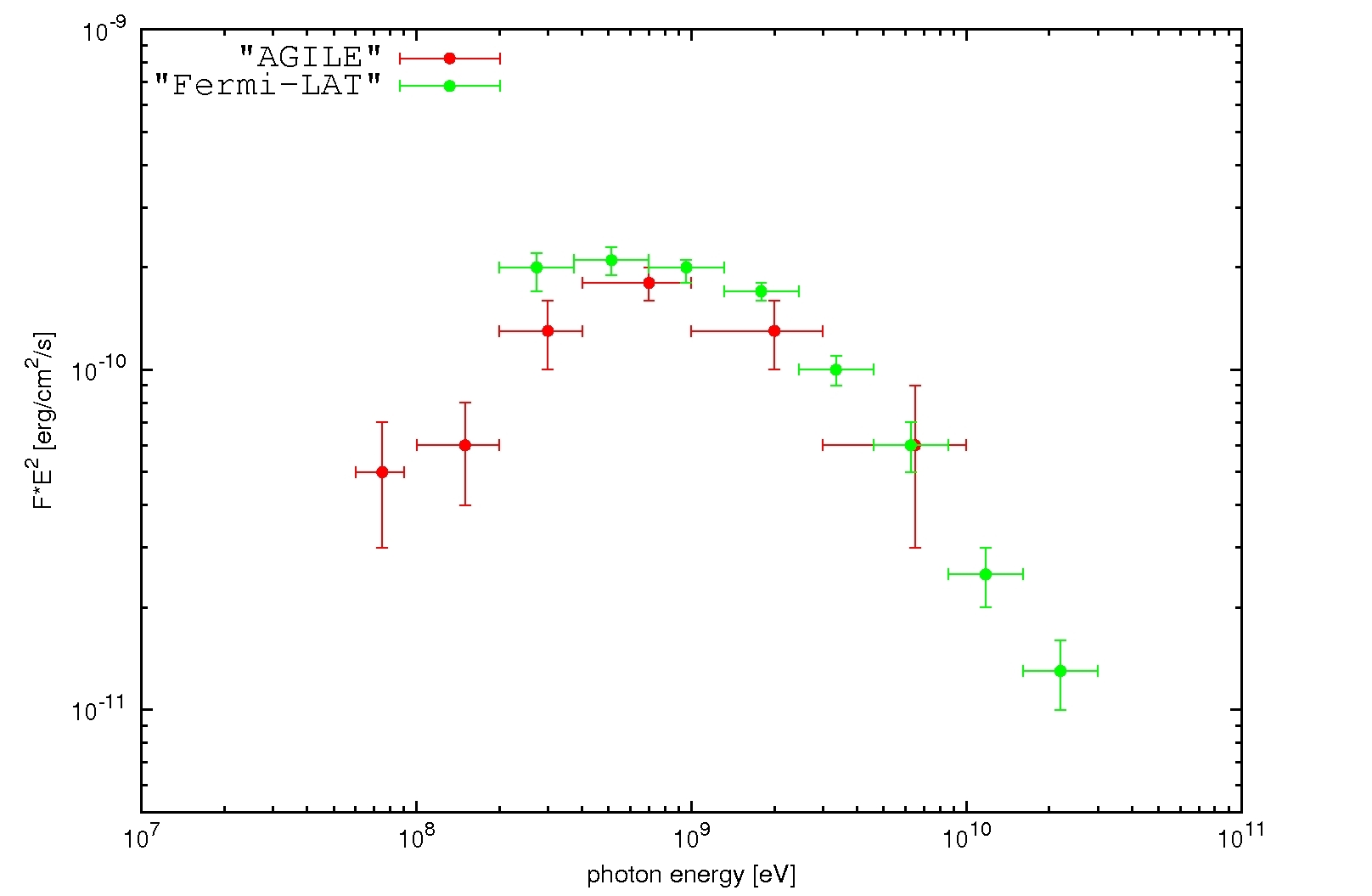}
\end{center}

\caption{Combined AGILE (red) and Fermi/LAT (green) spectral energy distribution (SED) for SNR~W44.  
AGILE points are in the range 50~MeV- 10~GeV divided into six energy intervals.
Fermi/LAT data span the energy range 0.2 - 30~GeV (from Abdo et al., 2010). }
\label{w44_spec}
\end{figure}

\subsection{The gamma-ray spectrum}
Figure~\ref{w44_spec} shows the AGILE W44 photon energy spectrum for the whole range 50~MeV- 10~GeV. 
The measured flux above 400~MeV is $F=(16.0\pm1.2)\times10^{-8}$~photons~cm$^{-2}$~s$^{-1}$.
In this figure are also shown Fermi/LAT spectral points \citep{abdo10W44} over an energy range 0.2-30~GeV. 
In the band where the spectra overlap, the sets of data are compatible at 1$\sigma$.
In this regard it is important to stress how AGILE is able to detect the emission from W44 in the energy range 50~MeV-300~MeV extending the spectrum to energies substantially lower than those previously obtained.
This spectrum shows a clear decrement at photon energies lower than 400~MeV, confirming the expectations based on neutral pion emission from accelerated protons/ions: a peak energy near 1~GeV for hadron energy spectra flatter than $E^{-2}$  \citep[e.g.,][]{aharoniana}. 
The gamma-ray spectrum becomes steep at higher energies with a photon power-law index $\alpha\sim 3\pm$0.1, in agreement with previous measurements \citep{abdo10W44}.

\subsection{Lepton-dominated models of emission}

\begin{figure}[h!]
 \begin{center}
 \includegraphics[width=9.3cm]{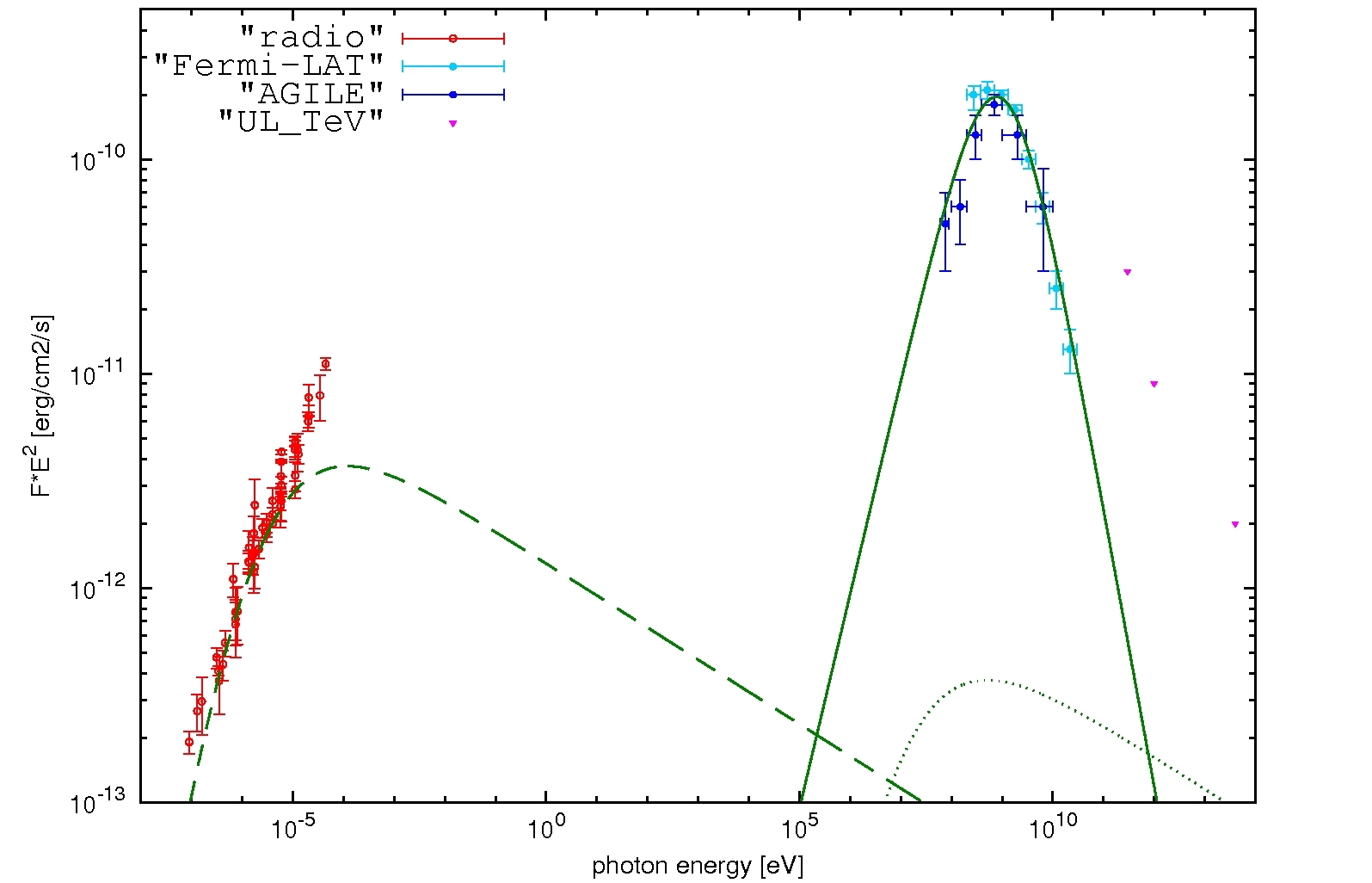}
\end{center}

\caption{Most favorable of the leptonic-only models, characterized by \textit{B}=20~$\mu$G and \textit{n}= 300~cm$^{-3}$.
The green curves show the electron synchrotron (dashed curve), Bremsstrahlung (solid curve), and IC (dotted curve) contributions.
This model is in contradiction with the radio data, and requires an average value of the gas number density that is too large compared
with the value deduced by radio and optical observations ($n\sim 100$~cm$^{-3}$). }
\label{spec3}
\end{figure}

Leptonic-only models of gamma-ray production have to satisfy the very well determined spatial and spectral constraints provided by the radio, optical,
and gamma-ray emissions. 
For both the Bremsstrahlung and inverse Compton cases we tested the class of leptonic-only models attempting to reproduce the observed gamma-ray spectrum
for different input leptonic spectra (see Table 1). 
{ We used a phenomenological approach based on two evidences: a) the synchrotron spectrum implies that the radio electron distribution
is well described by a power-law over a wide range of energies,  b) the discontinuity in the slope of the gamma-rays spectrum implies a discontinuity
in the emitting particle spectrum. }
{We assumed three electron distributions that can, in principle, describe this behavior: 1) a power-law with a high energy cut-off, 
$ F_1e (E) = K_e E^{-p} e^{-\frac{E}{E_c}}$,
\citep[as in][]{hendrick01}, 2) a power-law with a low energy cut-off, $F_2e (E) = K_e E^{-p} e^{-\frac{E_c}{E}}$, \citep[as in][]{gabici},
3) a broken power-law, $F_3e (E) = K_e \left(   \frac{E}{E_c}    \right)^{p_1} \left(    \frac{1}{2} \left( 1 + \frac{E}{E_c}  \right)  \right)^{p_1 - p_2} $,
\citep[as in][]{ahazira07}},
{where $E_c$ is the cut-off energy and $K_e$ is the normalization constant.}\\
{Since the distributions with a cut-off (1 and 2) fail to reproduce simultaneously both the radio and $\gamma$-ray spectrum, we refer to the distribution (3).}
{The best fit to the gamma-ray data is obtained with} $E_{c}=1$~GeV, and indices $p_{1}=0$ and $p_{2}=3.3$ above and below $E_{c}$, respectively. 
Synchrotron emission  originating from this distribution can be evaluated for different values of the average magnetic field. 
Figure~\ref{spec3} shows the case of the most ``favorable'' leptonic-only model characterized by $B=20$~$\mu$G
(other cases turn out to be even less favorable). 
We find that, at high frequencies, the calculated synchrotron spectrum is in strong disagreement (factor larger than 4) with the radio emission produced co-spatially to the gamma-ray one \citep[][and Figure~\ref{w44}]{castelletti07}.
Furthermore, the inferred average density, \textit{n}= 300~cm$^{-3}$, is too large (by a factor of three) compared with the circumstellar medium
constraints \citep{reach}; a lower value for $n$ would be incompatible with the gamma-ray spectrum fitting. 
Similar or even stronger contradictions with the multiwavelength data apply to
other leptonic-only models that we systematically explored
for a large variety of parameters. 
\\In the case of inverse Compton dominated models, two sources of soft photons are available: the cosmic background radiation (CBR) and the
interstellar radiation field (ISRF).
{In the first case, a second peak in the gamma-ray spectrum is unavoidably expected with a peak energy $E_{max}\sim$ 1 TeV, in contradiction with the upper-limits obtained from TeV Cherenkov telescopes. 
In the case of interaction with ISRF, instead, the calculated synchrotron peak is not compatible with the radio continuum data for any reasonable value of the magnetic field in the SNR shell.}
We can then reliably exclude leptonic-only models of emission for SNR~W44.

\subsection{Hadron-dominated models of emission}

\begin{figure}[h!]
 \begin{center}
 \includegraphics[width=9.3cm]{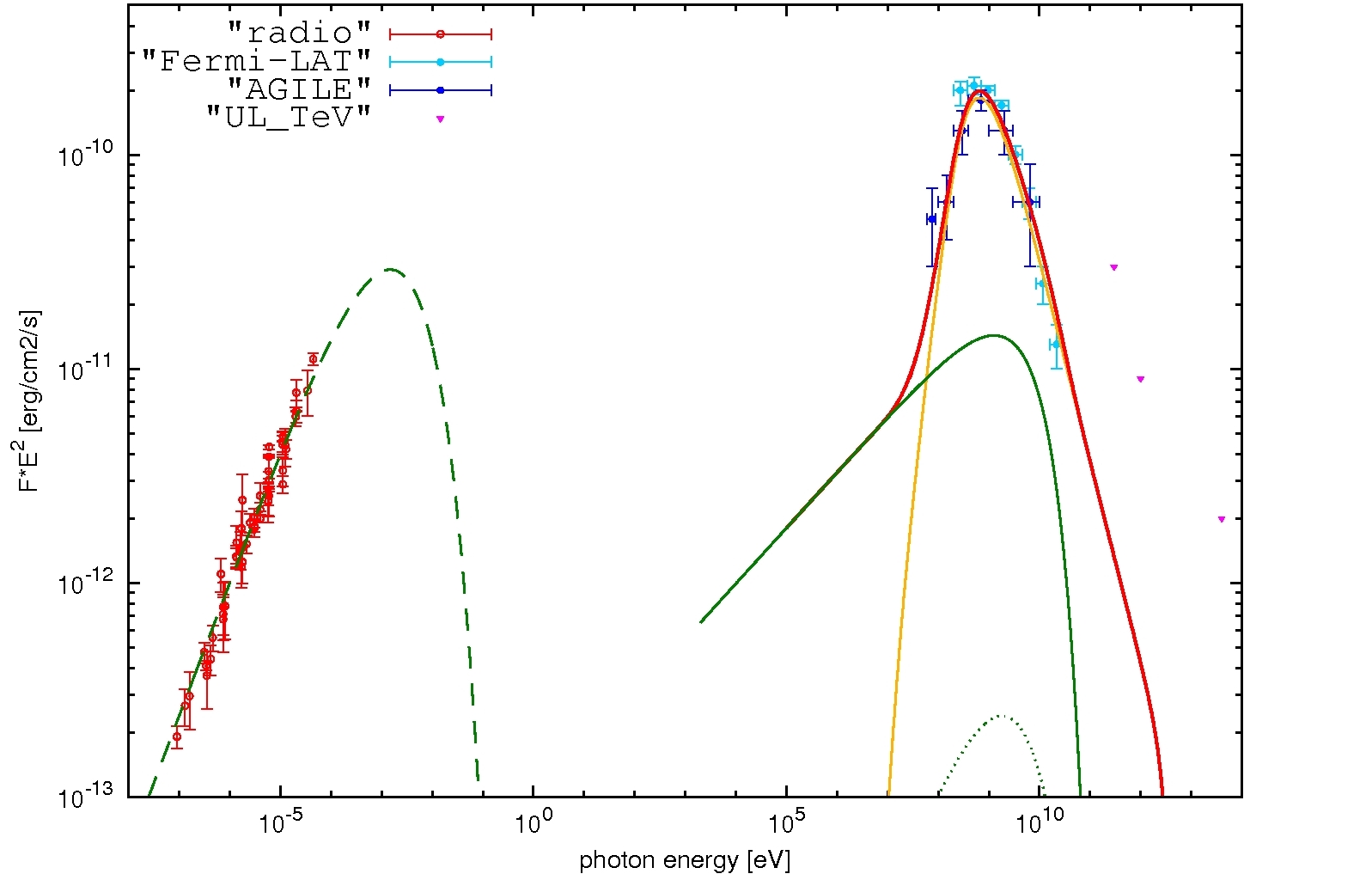}
\end{center}
\caption{Theoretical modeling of the broad-band spectrum of SNR W44 superimposed with the radio (data points in red color) and gamma-ray data of Figure~\ref{w44_spec} (in blue color).
 { TeV upper limits are also shown.}.
This is an hadronic model characterized by \textit{B} =70~$\mu$G and \textit{n} =100~cm$^{-3}$.
The yellow curve shows the neutral pion emission from the accelerated proton distribution discussed in the text.
The green curves show the electron contribution by synchrotron (dashed curve), Bremsstrahlung (solid curve), and IC (dotted curve) emissions.
The red curve shows the total gamma-ray emission. }
\label{spec2}
\end{figure}

It is interesting to determine the main physical parameters of an emission model dominated by hadrons in the gamma-ray energy range. 
{The gamma-ray emission was derived assuming that protons interact with the nuclei of the ambient medium through pp interactions
 and then radiate through $\pi^{0}$ decay \citep{kelner06}}.
{In order to fit the gamma-ray spectrum we tested, as proton energy distribution, a power law, a broken power law and a power law with cut-offs.  
The distribution providing the best-fit of the gamma-ray spectrum turned out} to  be a power-law with a relatively steep spectral index and a low-energy cut-off,
$F_p (E) \sim E^{-p} e^{-{E_{c}} / {E}}$, with $p=3.0\pm0.1$ and $E_{c}= 6\pm 1$~GeV. 
The inferred total energy of hadrons producing the observed gamma-ray spectrum is $E_{h}=10^{49}- 10^{50}$~erg (depending on the average magnetic field and
local densities, see Table 1) corresponding to a fraction of the total SNR shell kinetic energy $\epsilon\sim0.01-0.1$. 
{This value is a lower limit on the total energy of hadrons accelerated by W44 during its entire life, because the most energetic hadrons, likely accelerated during the early epochs of the SNR life, can escape from the acceleration site on timescales shorter than the age of this SNR \citep{berezinskii,gabici}.}
Electrons produce the radio continuum and shell-like features by synchrotron emission in an average magnetic field in the range $B\sim10-100$~$\mu$G. 
They also produce Bremsstrahlung and inverse Compton components that are subdominant in the gamma-ray energy range for a standard value of the electron/proton number ratio $\chi\sim0.01$. 
Our best ``hadronic model'' is shown in Figure~\ref{spec2}.
{We note that the spectral index found in this analysis is quite steep in comparison with the expectations for the spectrum of particles accelerated in SNRs \citep[e.g.][]{reynolds} and that the cosmic-ray spectrum can be altered (softened) by the interaction with sites where the target gas is confined, for example by molecular clouds  \citep[see e.g.][]{gabici10, ohira11}.
A comprehensive theoretical interpretation of this fact is beyond the scope of this work and will be addressed in a forthcoming paper.}

\section{Conclusions}

W44 turns out to be  an extremely interesting SNR whose environment and shocked material configuration favor a detailed
testing of hadronic versus leptonic emission. 
We find that models of accelerated protons/ions interacting with nearby dense gas successfully explain all the observed multifrequency properties of W44. 
The total number of accelerated hadrons in W44 is consistent with an efficiency of a few percent in the conversion of SNR kinetic energy into cosmic-ray energy. 

Within the hadronic scenario, the broad-band spectrum is successfully modeled by the gamma-ray emission resulting from the decay of neutral pions produced by accelerated protons/ions with a spectral index near 3 and a cut-off energy $E_{c}\sim6$~GeV (Figures~\ref{w44_spec} and \ref{spec2}).
Electrons (a few percent in number) contribute to the synchrotron radio emission and to a weak Bremsstrahlung component.
It is interesting to note that the spectral cut-off energy $E_{c}$ and the steep power-law index may indicate either a ``diffusion'' effect of high-energy hadrons, or a suppression of efficient particle acceleration in  dense environments \citep{uchiyama,malkov} as expected in W44.
Our results on W44 are complementary with those obtained in other SNRs, most notably IC~443 \citep{tavani} and W28 \citep{giuliani}.
Also in these SNRs the complex interaction of cosmic-rays with their dense surroundings produces gamma-ray spectra with high-energy cut-offs observed in the range 1-10~GeV.

\begin{acknowledgments}
We acknowledge stimulating discussions with F. Aharonian.
The AGILE Mission is funded by the Italian Space Agency (ASI) with scientific and programmatic participation by the Italian Institute
of Astrophysics (INAF) and the Italian Institute of Nuclear Physics (INFN).
Investigation carried out with partial support by the ASI grant no. {I/042/10/0}. 
G.D. and G.C. are members of CIC-CONICET (Argentina) and are supported by CONICET, ANPCyT and UBACYT grants.
\end{acknowledgments}

\newpage
\begin{center}
\textbf{Table 1}: Electronic component of hadronic and "ad hoc" leptonic models
\end{center}
\begin{table}[!h]
\footnotesize
\label{table}
\begin{tabular}{|l|c|c|c|c|c|c|p{6.5cm}|}
\hline
\textbf{Model}&$\mathbf{<n>}$& $\mathbf{<B>}$    & $\mathbf{E_{h}}$ & $\mathbf{E_{e}}$& $\mathbf{p_{1}}$   & $\mathbf{p_{2}}$    & \textbf{Comments} \\
&$\mathbf{(cm^{-3})}$  & $\mathbf{(\mu G)}$ & \textbf{(erg)} &  \textbf{(erg)}  &        &                 &        \\
\hline
&            &                            &                          &                           &             &       &$\bullet$ good agreement with data\\
& 20         &            20          &        $2.1\times10^{50}$  &   $2.0\times10^{49}$        &    1.7       &   3.0   &$\bullet$ \textit{n} small\\
Hadronic&            &                            &                           &                            &                 &         &$\bullet$ $E_{h}$ large\\\cline{2-8}
&     &                            &                          &                           &             &       &\\
&   100         &            70                &    $3.3\times10^{49}$ &    $2.8\times10^{48}$          &    1.7    &3.0    &   good agreement with all data\\
&     &                            &                          &                           &             &       &\\
\hline
         &                  &                    & &                                 &            &               & $\bullet$ incompatible with the spectrum\\
&      300          &          20       & - &       $7.3\times10^{48}$       &     0      &    3.3         &  $\bullet$ too large medium density\\
          &                &                     &    &                                &             & 	      & $\bullet$ too large $E_{e}$\\
 \cline{2-8}
Leptonic         &	            &                     &      &                            &         &            & $\bullet$ incompatible with the radio spectrum: wrong slope, low-frequency excess\\
               &             &                    &     &                                &         &          & $\bullet$ too large medium density\\
 &    5000          &      200            &  -   &       $4.4\times10^{47}$     &    0     & 3.3      &$\bullet$ too large average \textit{B}\\
           &                           &          &                 &                           &               &           &  $\bullet$ electron cooling (Bremsstrahlung) timescale too short with respect to the SNR age\\
\hline
\end{tabular}
\end{table}

\newpage

\end{document}